\documentclass[conference]{IEEEtran}
\IEEEoverridecommandlockouts
\usepackage{cite}
\usepackage{amsmath,amssymb,amsfonts}
\usepackage{algorithmic}
\usepackage{graphicx}
\usepackage{textcomp}
\usepackage{xcolor}
\usepackage{hyperref, url}
\usepackage{listings}
\usepackage{enumitem}\setlist[description]{font=\textendash\enskip\scshape\bfseries}
\usepackage[ruled,vlined]{algorithm2e}
\usepackage{chapterbib}
\def\BibTeX{{\rm B\kern-.05em{\sc i\kern-.025em b}\kern-.08em
    T\kern-.1667em\lower.7ex\hbox{E}\kern-.125emX}}
\begin{document}

\title{Using Fuzzy Matching of Queries to optimize Database workloads\\
}

\author{\IEEEauthorblockN{Sweta Singh}
\IEEEauthorblockA{
sweta\_singh@in.ibm.com}
\and
\IEEEauthorblockN{Vaibhav Kulkarni}
\IEEEauthorblockA{
vaibhav.kulkarni@in.ibm.com}

\and
\IEEEauthorblockN{Mario Briggs}
\IEEEauthorblockA{
mario.briggs@in.ibm.com}
\and
\IEEEauthorblockN{ Deepak Mahajan}
\IEEEauthorblockA{
deepakmahajan315@gmail.com}
\and
\IEEEauthorblockN{            Eitan Farchi }
\IEEEauthorblockA{
       farchi@il.ibm.com}
}

\maketitle

\begin{abstract}
Directed Acyclic Graphs (DAGs) are commonly used in Databases and Big Data computational engines like Apache Spark for representing the execution plan of queries. We refer to such graphs as Query Directed Acyclic Graphs (QDAGs). This paper uses similarity hashing to arrive at a fingerprint such that the fingerprint embodies the compute requirements of the query for QDAGs. The fingerprint, thus obtained, can be used to predict the runtime behaviour of a query based on queries executed in the past having similar QDAGs. We discuss two approaches to arrive at a  fingerprint, their pros and cons and how aspects of both approaches can be combined to improve the predictions. Using a hybrid approach, we demonstrate that we are able to predict runtime behaviour of a QDAG with more than 80\% accuracy.
\end{abstract}


\section{Introduction}
Workload management systems in a database or data processing engine intelligently schedule incoming requests or queries to ensure optimal utilization of resources. They act as a gatekeeper of the system, admitting optimal number of queries at a time, such that the stability and performance goals can be met. Queries that cannot be admitted, because they will affect the stability and performance of the system, are queued up for execution at later point in time. Towards achieving the stability and performance goals, such systems classify an incoming query based on its complexity. The execution time of a query is directly proportional to its resource usage and can serve as a measure of complexity. Workload Management systems can maintain a separate lane for each category, like Simple, Medium and Complex, each of which has its share of system resources. These lanes can then be managed using different policies, such that the performance goals for each lane can be met. For example, for a Simple lane consisting of relatively short duration queries, fast response time is very critical. Therefore, queries in a Simple lane can be ordered based on their estimated execution time. Furthermore, a query in Simple lane may not need lot of CPU cycles for execution. So, Simple lane can be allocated lower CPU resource share. On the other hand, a complex query runs relatively longer and is resource hungry. Hence, queries in a Complex lane could benefit from a higher share of CPU resources. Thus, predicting the complexity of a query accurately is core to building intelligent Workload Management Systems.

Data processing engines like Apache Spark use DAGs (Directed Acyclic Graphs) to represent execution plans for queries. DAG consists of vertices (or nodes) and edges, with each edge directed from one vertex to another, such that following those directions will never form a closed loop. We refer to DAG which is used to represent query execution plan as QDAG. The vertices (or nodes) represent operators like Scan, Filter, Join, Aggregation. Structure of QDAG represents order of execution and dependency between these operators. Usually, the complexity of queries depend on the number and type of vertices and edges in the QDAG. Higher the number of vertices and edges in a QDAG, more complex is execution of a query. Workload Management Systems can understand the complexity of the incoming query by finding a similar query from previously executed queries. This requires a way to compare the similarity between QDAGs of the queries. One way to measure the similarity between QDAGs could be to learn vector embedding of a plan, \cite{query2vec}. Similarity between two queries can then be expressed as distance between their vector embedding. However, this method has its own set of challenges. It requires a training phase and availability of labelled dataset.
In this paper, we introduce the notion of similarity for QDAGs using Moses S. Charikar's Similarity estimation technique, \cite{simhash}. Charikar's simhash, \cite{simhash} is a dimensionality reduction technique. It maps high-dimensional vectors to small-sized fingerprints. 
\\
\\
While similarity hashing has been used for finding similar documents and detecting vulnerabilities in binaries, to the best of out knowledge, ours is the first attempt to fingerprint QDAGs. Our key contributions in this paper are as follows: \\
1. Generating a compact 128-bit fingerprint of a QDAG, such that similar QDAGs will have similar fingerprints. \\
2. Proposing two alternate ways to generate a QDAG's fingerprints using similarity hashing, \cite{simhash}, their pros and cons, and recommendations on which method to use based on different scenarios. \\
3. Demonstrating that this technique is able to predict a query’s complexity with around 80 percent accuracy. \\
\\
Workload management system can utilize this concept of QDAG fingerprints to understand the complexity of incoming queries. Workload management system can maintain an in-memory lookup table that stores QDAG fingerprint of past query execution instances, along with their execution time. When a new query arrives, fingerprint of QDAG can be matched to fingerprints of QDAGs stored in the in-memory lookup table. The matched query's complexity can be used to predict the incoming query's complexity. 
\\
\\
Rest of the paper is organized as follows. Section II introduces the concept of applying Simhash to QDAGs. Section III provides details of the two alternate ways of generating QDAG fingerprints. Section IV provides details of the test workloads used for the experiments. 
Section V provides details of how the QDAG fingerprints can be used to predict query runtime complexity. Section VI provides prediction accuracy and a general evaluation of the two fingerprinting techniques discussed in Section IV. Section VII discusses the pros and cons of the two fingerprinting techniques and Section VIII provides conclusion on the two fingerprinting techniques \\

\section{Fingerprinting with Simhash}
Simhash possesses two properties: 
(A) The fingerprint of a QDAG is a “hash” of features of all the nodes, and 
(B) Similar QDAGs have similar hash values. 
The latter property is atypical of hash-functions. For illustration, consider two QDAGs that differ in a very small way. Consider two queries with similar QDAGs except for different column names in the project operation. Then hash functions like  Cityhash, \cite{cityhash} will hash these two QDAGs (treated as strings) into two completely different hash-values (the Hamming distance between the hash values would be large). However, Simhash will hash them into similar hash-values (the Hamming distance would be small).
\\
\\
Let V represent the n-bit final fingerprint. Each node and its properties in a QDAG is hashed into an n-bit hash value.
These n bits increment/decrement the n components of the vector, V, as follows: 
if the i-th bit of the hash value is 1, the i-th component of V is incremented by 1;
if the i-th bit of the hash value is 0, the i-th component of V is decremented by 1. 
When all nodes have been processed, some components of V are positive while others are negative. The signs of components determine the corresponding bits of the final fingerprint, V.
\\

\section{Fingerprinting QDAGs}
We generate two fingerprints for a QDAG, one for all the nodes and it's features and second for the edges. We fingerprint the edge properties,  \ref{fingerprintedges} in 64-bit vector using the method described in subsection below,  \ref{SummariseEdges}. To fingerprint the nodes, we experimented with the following two approaches.\\
\\
Approach \#1: The node's fingerprint is arrived at by carefully engineering each property of the node using positional encoding. Most of the properties are kept schema independent, \ref{Glossary}. Examples of properties are number of string and numeric attributes in grouping expressions, result expressions, keys of join and project operator, join type, partitioning type, mode of BroadcastExchange \ref{Glossary}, etc. The encoding of a node, thus obtained, is hashed to 64 bits vector. Hash of all nodes are combined using simhash.\\
\\
Approach \#2: The node's fingerprint is computed by treating each node and its properties as a sequence of characters, breaking it up into n-grams, hashing the n-grams of all nodes and finally, combining them using simhash. \\
\\
For both approaches, the algorithm for fingerprinting edges remains same. It is described in the following section

\subsection{Summarising edges of QDAG}
\label{SummariseEdges}
\begin{figure*}[h!]
\centerline{\includegraphics[scale=0.52]{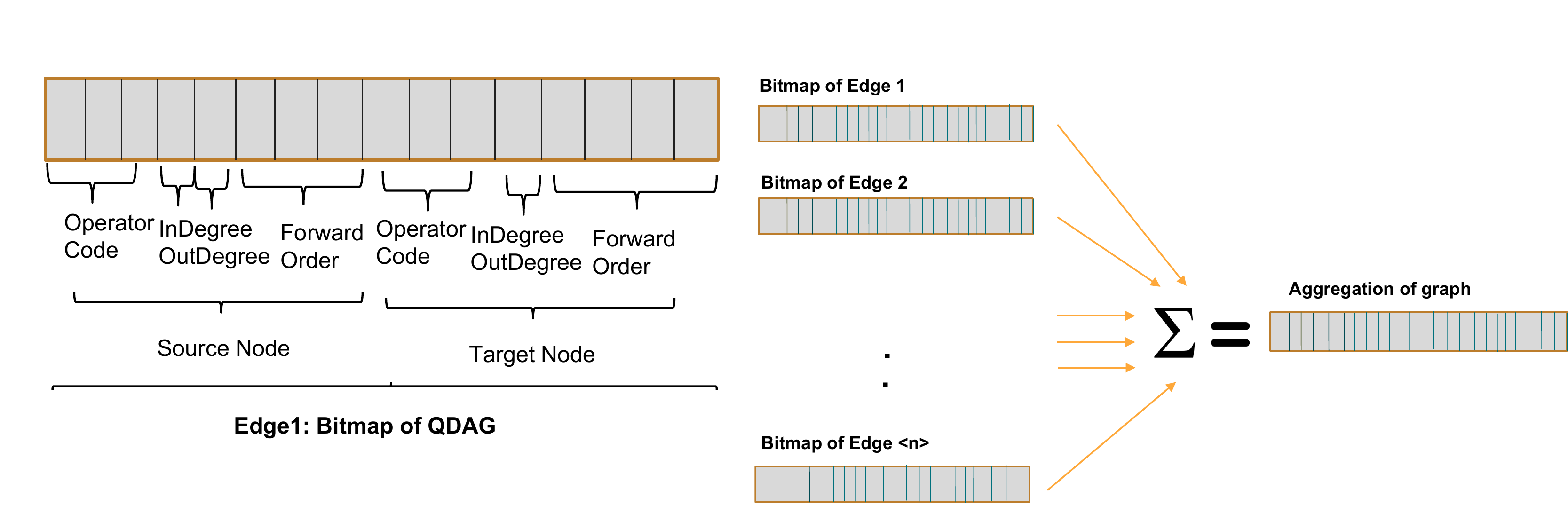}}
\caption{Fingerprinting edges}
\label{fingerprintedges}
\end{figure*}

For each edge in the QDAG, we find out the forward topological order, \ref{Glossary}, backward topological order, \ref{Glossary}, in-degree, \ref{Glossary}, out-degree, \ref{Glossary} and operator of the connecting source and target nodes. After computing these properties, we generate a bitmap vector of an edge using positional encoding. In this method, every bit in the vector represents an edge attribute as depicted in Figure \ref{fingerprintedges}. We add the vector of all the edges to arrive at the final representation of the QDAG. The pseudo code for the hash calculation can be found in Algorithm, \ref{algorithm1}. The shift left function in the algorithm shifts the bits by a constant number of positions to the left. \\

\begin{algorithm}
\SetKwInOut{Input}{inputs}
\SetAlgoLined
\KwResult{64 bit representation of the edges S(E) }
 \Input{QDAG G = (V,E)}
 where V is Nodes in the QDAG and E are the edges connecting the nodes
 
 $G_r$ = reverse\_edges\_of\_QDAG(G) \; 
 forward\_order = topological\_sort(G)\; 
 backward\_order = topological\_sort($G_r$) \;
 hash = 0 \;

where u and v are nodes connected by an edge,
 \ForEach{(u,v) $\in$  E}{
  bitmapEdge = 0 \\
  
  temp = operator\_code[u] ; \\
  bitmapEdge += shift\_left(temp) ; \\ 
  temp = forward\_order[u] ; \\
  bitmapEdge += shift\_left(temp) ; \\ 
  temp = backward\_order[u] ; \\
  bitmapEdge += shift\_left(temp) ; \\ 
  temp = in-degree[u]; \\
  bitmapEdge += shift\_left(temp) ; \\
  temp = out-degree[u];  \\
  bitmapEdge += shift\_left(temp) ; \\
  
  temp = operator\_code[v] ; \\
  bitmapEdge += shift\_left(temp) ; \\ 
  temp = forward\_order[v]; \\
  bitmapEdge += shift\_left(temp) ; \\
  temp = backward\_order[v]; \\
  bitmapEdge += shift\_left(temp) ; \\
  temp =  in-degree[v]; \\
  bitmapEdge += shift\_left(temp) ; \\
  temp = out-degree[v]; \\
  bitmapEdge += shift\_left(temp) ; \\
  
  bitMapEdges += bitMapEdge; \\
 }
 S(G) $\gets$ bitMapEdges;
 \label{algorithm1}
 \caption{Computation of Structural representation of the graph}
\end{algorithm}

\vspace{0.5cm}
\subsection{Approach \#1 Details: Computation of the node fingerprint}
\begin{figure*}[h!]
\centerline{\includegraphics[scale=0.52]{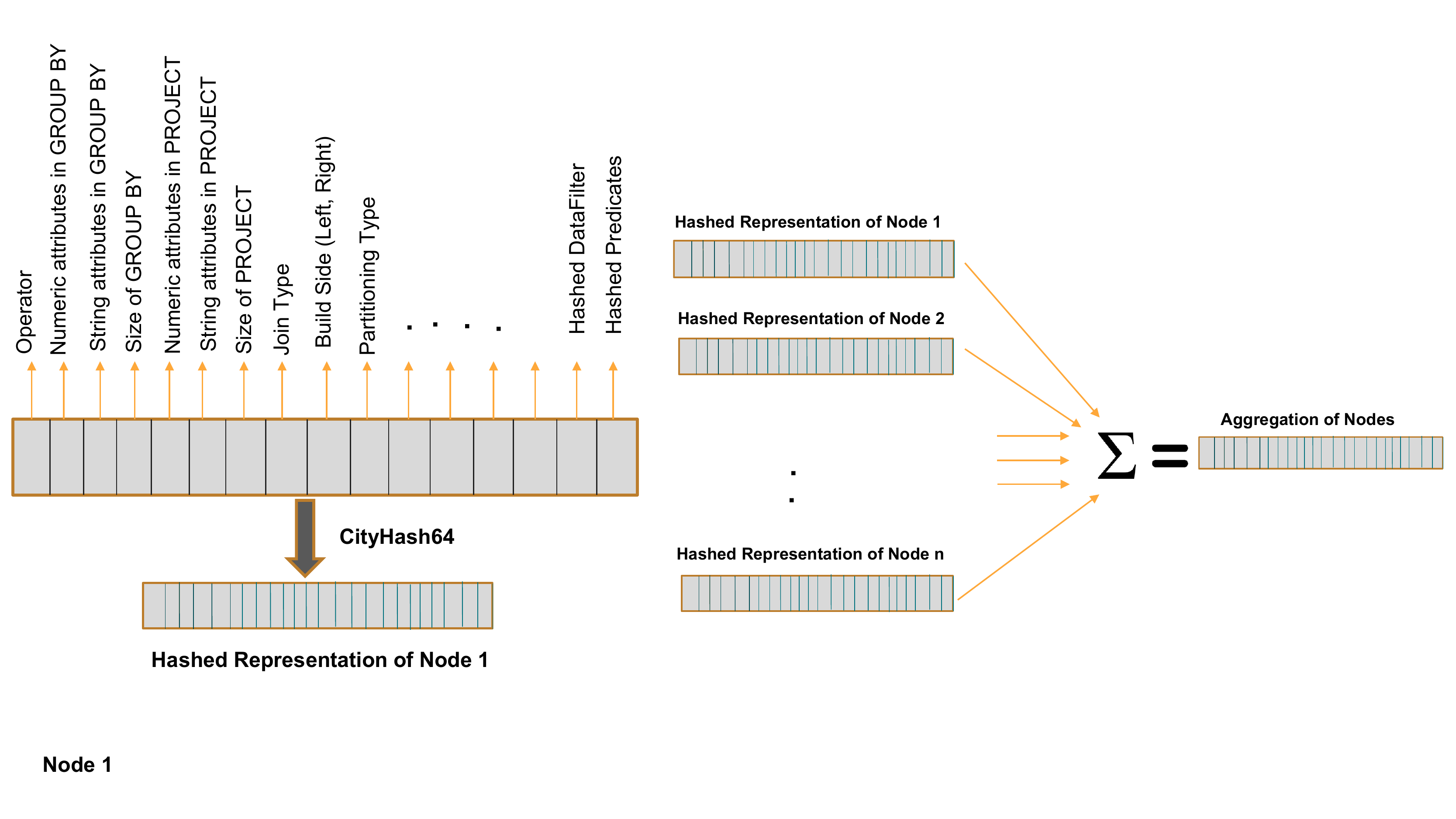}}
\caption{Approach \#1 Fingerprinting nodes}
\label{figureapproach1fingerprintnode}
\end{figure*}

In this approach, we perform feature engineering of the operators in the QDAG. After parsing and extracting the node features, we get a vector of features for each node in the QDAG as shown in Figure \ref{figureapproach1fingerprintnode}. For node $v \in V$ we represent the set of features by f(v). After getting f(V) for each node v, we convert f(v) into a string and then hash it using Cityhash, \cite{cityhash} as shown in Algorithm,  \ref{nodefingerprintapproach1}. Lets denote this hash of the node v by h(v). This acts as a representation of the node features for the node v. After getting h(v), a representation for each node in the QDAG, we need an aggregation method in order to summarise all nodes into a single hash. Algorithm for summarizing is shown in Algorithm,  \ref{combinenodefingerprintsapproach1}. This final hash will act as a compact Fingerprint of the whole QDAG. 
\begin{algorithm}
\SetKwInOut{Input}{inputs}
\SetAlgoLined
\KwResult{64 bit representation for each node h(v)  }
 \Input{Node v $\in$ V }   

  feature\_vector $\gets$ [];\\
  feature\_vector.add(encoded\_operator(v) );\\

 \ForEach{property p $\in$ P}{
  \algorithmicif{~property p in v} 
  \algorithmicthen{~feature\_vector.add(p)}\\
 }

 \textbf{return} CityHash64( string( feature\_vector ) )
 \label{nodefingerprintapproach1}
 \caption{Algorithm for computing fingerprint of single node in Approach \#1}
\end{algorithm}

\begin{algorithm}
\SetKwInOut{Input}{inputs}
\SetAlgoLined
\KwResult{64 bit representation of the graph features N(G) }
 \Input{QDAG G = (V,E)}   
 \ForEach{operator v $\in$  V}{
  hash(v) $\gets$ vectorize( h(v) ) \\
  weighted(v) $\gets$ depth(v) $\times$ hashV \\
  }
  \ForEach{bit in range(64)}{
    sum $\gets$ 0\\
    \ForEach{v $\in$ V}{
        sum $\gets$ sum + weighted(v)[bit] \\
    }
    
  \algorithmicif{sum > 0} 
  \algorithmicthen{SimHashG[bit] $\gets$ 1}\\
  \textbf{else} SimHashG[bit] $\gets$ 0  \\

 }
 \textbf{return} SimHash
 \label{combinenodefingerprintsapproach1}
 \caption{Algorithm for summarizing fingerprints of all nodes in QDAG using simhash in Approach \#1}
\end{algorithm}

\subsubsection{Generating Node Fingerprint using simhash}
\label{simhash}
For summarising the node features of the whole graph we use the simhash algorithm introduced by Charikar, \cite{simhash}. So, given the hashes of all the nodes in the graph (h(v) $\forall v \in V$), we calculate Simhash(V) that is a 64 bit representation of the whole graph G. Let the $i^{th}$ bit of h(v) be $h(v)_i \forall i \in [1,64]$. Now, for weights $W_v \forall v \in V$ we find bitcount for all bit positions and represent it by an array BitCount. Therefore, $BitCount_i = \sum_{v \in V, h(v)_i = 1}W_v - \sum_{v \in V, h(v)_i = 0}W_v$ and the final hash of the graph can be found by checking which positions in BitCounts are positive. \\

\begin{equation}
Simhash(G)_i =
	\begin{array}{ll}
		1  & \mbox{if } BitCount_i > 0 \\
		0 & \mbox{if } BitCount_i \leq 0
	\end{array}
\end{equation}

This gives us the final 64 bit representation for the node features of the graph. So N(G) = Simhash(G) 

\begin{figure*}[h!]
\centerline{\includegraphics[scale=0.52]{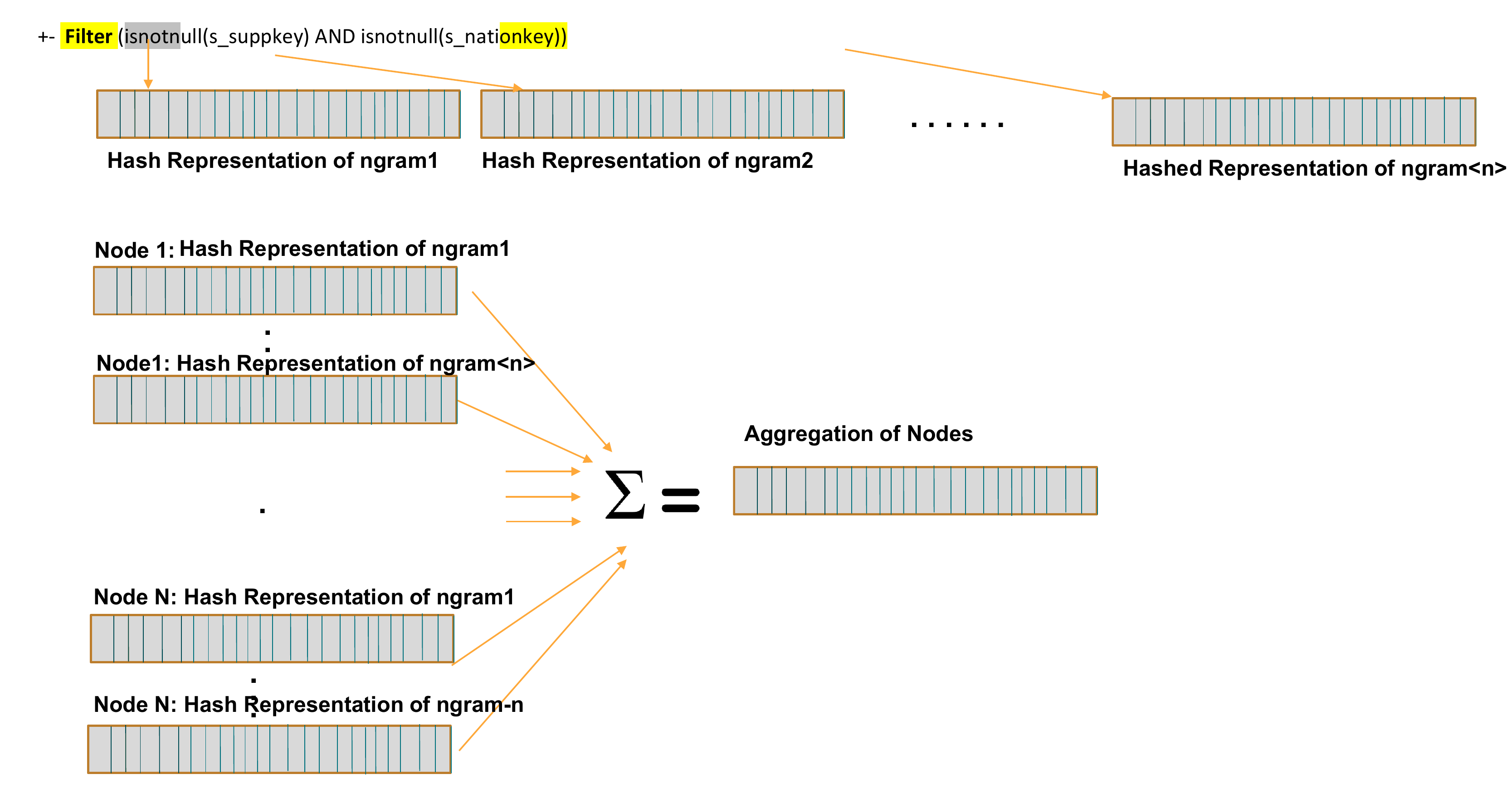}}
\caption{Approach \#2 Fingerprinting nodes}
\label{ngramtsne3}
\end{figure*}

\vspace{0.5cm}
\subsection{Approach \#2 Details: Computation of the node fingerprint}
Following the work of Gascon et.al.\cite{ngram}, we present another method for generation of a fingerprint for the QDAG. In this method, the QDAG is seen as a stream of characters, and n-gram on operators and their properties are computed. All n-grams are hashed using non-cryptographic hashing functions and combined using simhash. For each node of the graph, we have a line in QDAG representing the query operator and attributes or arguments to the operator in the form of string. For each node $v \in V$, we treat the whole string consisting of the attributes and operator name as a fact. We calculate n-grams of the fact and hash every ngram to get a set of hashes for each node. 

$$hashes_{node}(v) = \{ CityHash(g) | \forall g \in N-Grams(fact(v)) \}$$
where N-Grams(fact(v)) represents the set of all N grams of string of node v. In our experiments, we used 3-grams to get the set of hashes for every node. After getting the set of hashes for each node, we combine the hashes using simhash to get a fingerprint for all nodes in the QDAG

\begin{algorithm}
\SetKwInOut{Input}{inputs}
\SetAlgoLined
\KwResult{64 bit representation of the node features N(G) }
 \Input{direct acyclic graph G = (V,E)}  
 node\_simhashes $\gets$ []; \\ 
 \ForEach{operator v $\in$  V}{
  n-grams(v) $\gets$ FindAllNGrams(v) \\
  Hashes(v) $\gets$ [] \\
  \ForEach{ng $\in$ n-grams(v)}{
    Hashes(v).append(vectorize( CityHash(ng))); \\
  }
  node\_hashes.append(Hashes(v)) ; \\ 
 }
 FinalHash $\gets$ SimHash( node\_hashes ); \\
 \textbf{return} FinalHash
 \label{approach2nodefingerprint}
 \caption{Algorithm for computing fingerprint of nodes in QDAG in Approach \#2}
\end{algorithm}

\begin{algorithm}
\SetKwInOut{Input}{inputs}
\SetAlgoLined
\KwResult{ SimHash(H) }
 \Input{vector of hashes H in which every element is a vector of 64 bits }   

  \ForEach{bit in range(64)}{
    sum $\gets$ 0\\
    \ForEach{h $\in$ H}{
        sum $\gets$ sum + h[bit] \\
    }
    
  \algorithmicif{sum > 0} 
  \algorithmicthen{SimHash[bit] $\gets$ 1}\\
  \textbf{else} SimHash[bit] $\gets$ 0  \\

 }
 \textbf{return} SimHash
 \label{simhashfromvector}
 \caption{Algorithm for combining a vector of hashes to get a single 64 bit fingerprint using simhash}
\end{algorithm}

\section{Data}
\label{data}
We conducted our experiments on TPC-H \href{http://www.tpc.org/tpch/} and TPC-DS benchmarks \href{http://www.tpc.org/tpcds/} in Apache Spark. Apart from the standard 22 TPC-H and 99 TPC-DS queries, we added 10 and 100 additional queries to TPC-H and TPC-DS respectively. The additional queries provide representation for simple queries. 
Spark splits a query into stages, \cite{sparkDAG}. A stage is an independent unit of execution, and is computed in parallel by tasks. We generate fingerprints at the stage level for following reasons:
\begin{enumerate}
\item The fingerprint generation is faster, as the similarity hashing algorithm has to operate on a smaller sub-graph of a stage, instead that of a query.
\item The probability of finding similar fingerprints increase significantly, as QDAGs are very likely to share subgraphs like a table scan or joins between fact and dimension tables.
\end{enumerate}
In our experiments, we classify the QDAGs of TPC-DS, \cite{tpcds} and TPC-H, \cite{tpch} queries based on the complexity into Simple, Medium and Complex categories. QDAG with runtime less than 5 seconds is Simple, QDAGs with runtime more than 5 seconds and less than 30 seconds is Medium and QDAGS with runtime more than 30 seconds are Complex.

\section{Matching QDAGs using fingerprints}
\label{matching}
In order to perform the similarity matching for QDAGs, we compute and persist the fingerprints of the edges and nodes of QDAGs of past query executions in an in-memory hash map. When a new query hits the system, two 64-bit signatures for the edges(S($G_{key}$)) and nodes N($G_{key}$)   is computed and matched using the following two-step process:

\begin{enumerate}
\item First, we find a constant number of QDAGs that are similar to the given QDAG in terms of edges. For this, we calculate the hamming distance between S($G_{key}$) and S(G) $\forall G \in \mathcal{G}$ where $\mathcal{G}$ is the set of all the QDAGs in the system. After calculating the hamming distance, we filter out the QDAGs with the least hamming distance. Let the set of the QDAGs that we get after this filtering be $\mathcal{C}$. Note that $|\mathcal{C}| = k$(constant). 
\item Now, after getting the set $\mathcal{C}$, we find hamming distance between N($G_{key}$) and $N(G) \forall G \in \mathcal{C}$ and sort them in increasing order of the hamming distance. These are the nearest neighbours or the final matches for the given QDAG $G_{key}$. 
\end{enumerate}

\section{Evaluation}
For evaluating our method of matching, we predict the complexity of the QDAG for TPC-DS,  \cite{tpcds} and TPC-H,  \cite{tpch} queries.  In our experiments, we predict the complexity of a given QDAG by using the K-nearest Neighbors algorithm, \cite{knn}. Using the matching algorithm as described in section \ref{matching}, we find the nearest neighbor of the given QDAG using the it's fingerprint. The complexity of that nearest neighbor is the predicted complexity of the given QDAG. \\

\begin{table}[h]
\caption{Accuracy of QDAG complexity prediction}
\begin{center}
\begin{tabular}{|l|c|c|c|}
\hline
\cline{1-3} 
\textbf{} & \textbf{\textit{Approach \#1}}& \textbf{\textit{Approach \#2}}\\
\hline
Prediction Accuracy & 80.87\% & 82.72\%  \\ & & \\
\hline
Prediction Error & & \\ (Actual complexity is Simple, & & \\ Predicted is Medium or Complex) & 8.3\% & 6.4\% \\ & & \\
\hline
Prediction Error & & \\ (Actual complexity is Medium & & \\ or Complex, Predicted is Simple) & 17.5\% & 17.3\% \\ & & \\
\hline
\end{tabular}
\label{tab1}
\end{center}
\end{table}

Prediction Accuracy for both approaches is similar with Approach \#2 having a slight edge of 1.8\% over Approach \#1. We calculate the cost of mis-prediction in two cases viz. Simple QDAGs predicted as Medium or Complex and Medium or Complex QDAGs predicted as Simple. The first case can have higher cost on System throughput as Simple QDAGs might get starved of resources in a Medium or Complex lane. The second case may not have high cost on System throughput. Taking the first error case, we see that Approach \#2 performs better marginally by 1.9\% compared to Approach \#1. \\

In order to understand the effectiveness of both approaches and gain insights into their decisions, we decided to use SPSS Modeler's AutoClassifier. SPSS Modeler is an IBM software which provides variety of modelling methods taken from machine learning, artificial intelligence and statistics. SPSS AutoClassifier tests and compares various classification models in a single run and provides the models which are best suited for the given inputs. We fed the AutoClassifier with the predicted complexity from both approaches. The input features for the AutoClassifer are the nearest neighbour's node distance and the nearest neighbour's edge distance and the target variable is predicted complexity. For Approach\#1, the AutoClassifier tried different models on the input data to find out the model that suits the best for the input. The SPSS Autoclassifier suggested few models with prediction accuracy measuring from 81.14\% for CHAID model to 87\% for Random Forest model. For Approach \#2, SPSS AutoClassifier suggested models with prediction accuracy varying from 82.8\% for CHAID model to 84.56\% for Random Forest model. Since ensemble models like Random Forest or Gradient Boosted tress are hard to interpret, we used  CHAID model to explain the decisions taken by both approaches, Approach \#1 and Approach \#2.\\

\begin{figure*}[h!]
\centerline{\includegraphics[scale=0.20]{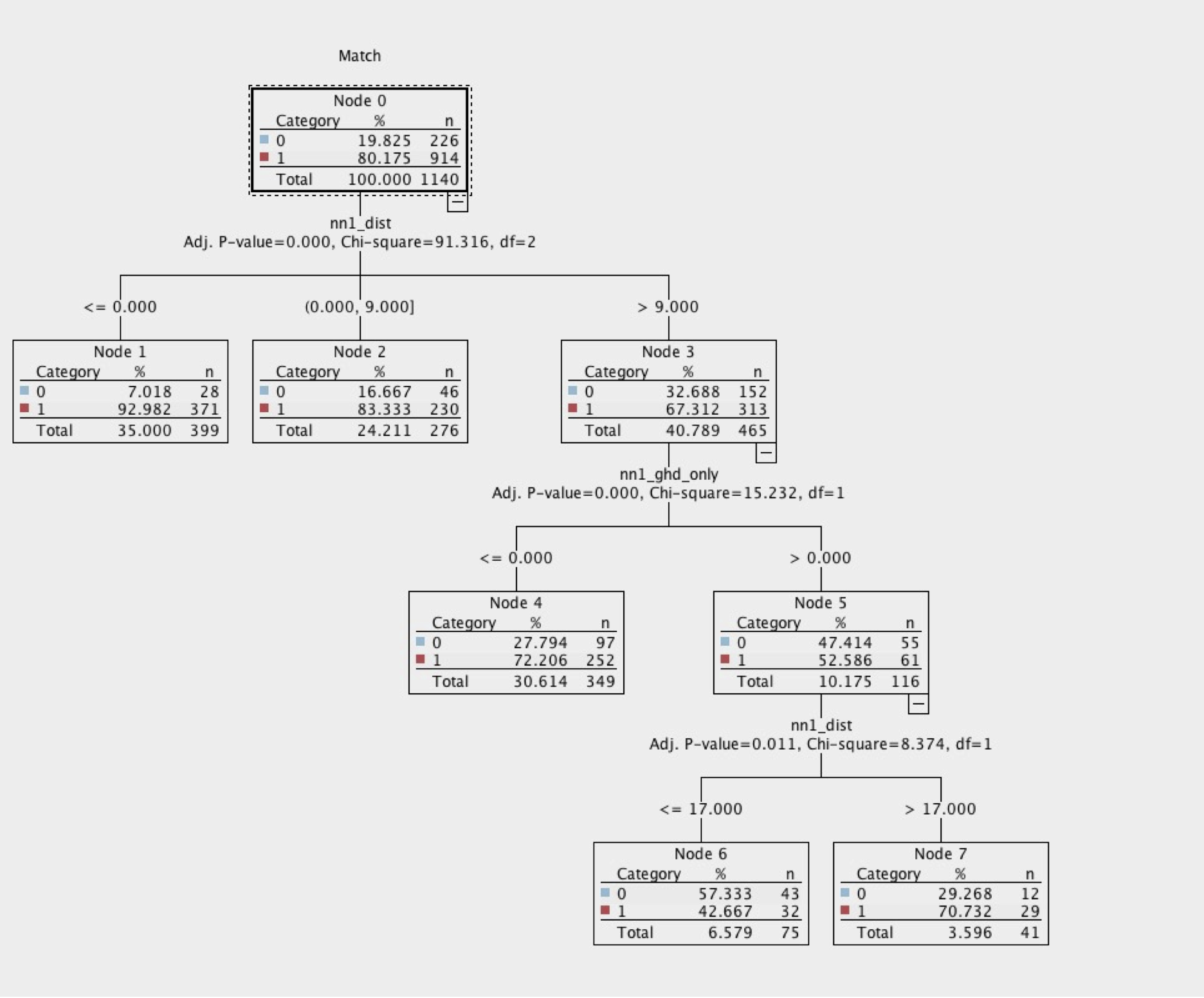}}
\caption{CHAID model decision tree for Approach1}
\label{approach1chaidmodel}
\end{figure*}

\begin{figure*}[h!]
\centerline{\includegraphics[scale=0.20]{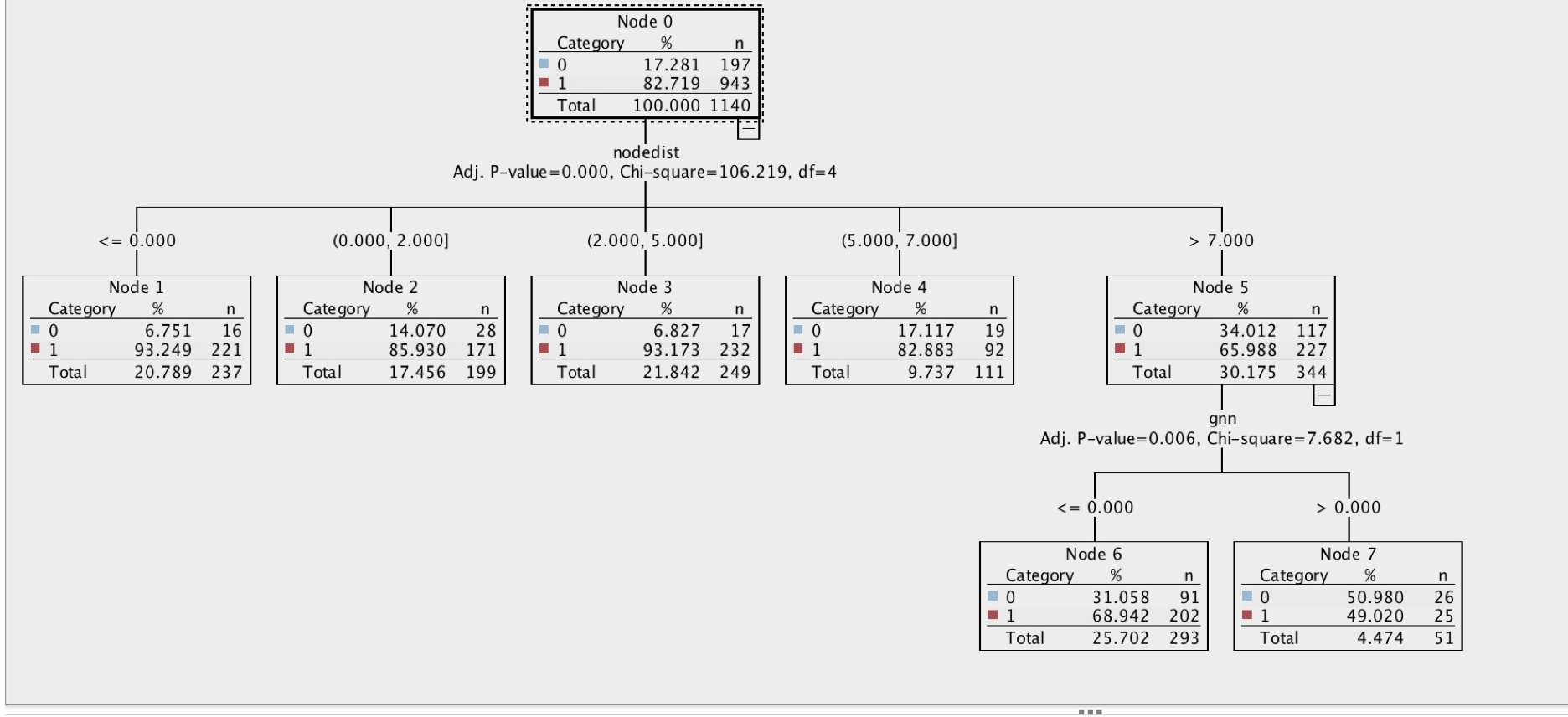}}
\caption{CHAID model decision tree for Approach2}
\label{approach2chaidmodel}
\end{figure*}

As seen in the Fig 4 for Approach \#1 CHAID model, as the "Nearest Neighbour node distance" increases, the prediction accuracy decreases. When the node distance is closer to 0.0, the prediction accuracy is 92.98\%, for node distances between 0.0 and 9.0, the prediction accuracy drops slightly to 83.33\%. For node distances greater than 9.0, the prediction accuracy depends on the "Nearest Neighbor Edge" distance. The prediction accuracy decreases with increased edge difference. This shows that if the QDAGs are such that their node distances is minimal, the complexity of the QDAGs can be predicted fairly accurately. Similarly, Fig 5. for Approach \#2, for node distances close to 0.0, the prediction accuracy is around 93.24\%. As the node distance increases the prediction accuracy decreases. For node distances between 0.0 and 2.0, the prediction accuracy drops to 85.93\%. Interestingly, for node distance between 2.0 and 5.0, the prediction accuracy increases to about 93.17\%. We believe that with more samples in the case where node distance is between 0.0 and 2.0, prediction accuracy of 85.93\% should improve and we would see prediction accuracy consistent with the node distance. i.e. as node distance is less, the prediction accuracy increases. Similarly, for node distances between 5.0 and 7.0, the prediction accuracy drops 82.88\% and so on. \\

Overall, the two approaches are able to predict QDAG's complexity with relatively higher accuracy when the closest matching QDAG has lower node and edge distance. Going by the overall model accuracy reported by the Auto-Classifier, Approach \#1 fares better than Approach \#2. This suggests that the models are able to find a consistent pattern in the predictions of Approach \#1, compared to Approach \#2. This reflects in the CHAID tree model too, where prediction accuracy decreases proportionately with increasing node and edge distance in Approach \#1. Approach \#2, on the other hand, is very sensitive to node distance with an anomaly in between where the prediction accuracy is at 85.93\% when node distance between 0.0 and 2.0, and improves when node distance is between 2.0 and 5.0.

\section{Discussion}
In the two approaches discussed in the paper, the first approach Approach \#1 has the following benefits:
\begin{enumerate}
\item Ability to hand-pick important features of operators that can be useful in predicting complexity. Features like the type of join ({Hash Join or Sort Merge Join or Broadcast Join}, {Inner, Outer, AntiJoin, SemiJoin}), which side of the join is the build side (left, right), the type of output partitioning ( hash or range or single partition), number of partitions are represented in the fingerprint.
\item Flexibility to engineer features in such a way that they are independent of the schema and hence, can be generalized. To achieve this, we avoid references to table names or column names. Instead, we use the data type of columns, number of numeric columns, number of string columns and width of rows.
\item Building a position based encoding where every position in the vector always represents a specific property of the operator makes this approach more interpretable. 
\item Ability to incorporate "insider" information, which is not very obvious to one not familiar with the workings of the Query optimizer. Example, in Apache Spark, there are special operators like ReusedExchangeExec or Subquery or ReusedSubquery. ReusedExchangeExec is an optimization that enables stages to reuse shuffle data from previous stage's execution. The subgraph of ReusedExchangeExec has to be looked up in the original stage's plan graph, reconstructed and encoded, ensuring that its semantics is not lost. Accurately representing such features improves the ability of algorithm to match such operators across different QDAGs and infer its complexity.\\

\end{enumerate}
Augmenting this approach with runtime statistics, like cardinality of each operator, can lead to a more generalized prediction model. Its predictions, as illustrated via machine learning models, follows a more consistent pattern. The downside of this approach is that it requires an understanding of QDAGs, the operators in the QDAGs, and their properties. In a way, this requires a deep understanding of how Database optimizer works. It also requires carefully selecting the important properties and features, encoding it in the vector representation. So, the engineering effort required is relatively high compared to Approach \#2. \\

The second approach Approach \#2, on the other hand, has the following benefits:
\begin{enumerate}
\item It is relatively simpler to compute the fingerprint. The engineering effort involved to construct a compact fingerprint is very low compared to Approach \#1.  One doesn't need expertise in Query optimizer and its internal workings.
\item Most data processing engines or databases compute hash treating the QDAG as a string, hence this method can easily replace the existing hash method.
\item This method can be applied to any database or data processing engine, without requiring any modifications.\\
\end{enumerate}

Compared to Approach \#1, the most appealing aspect of Approach \#2 is its simplicity and ease of implementation. It is very easy to have this approach up and running, with minimal pre-processing to eliminate elements that are very specific to QDAGs like operator identifiers. With very less effort, this approach fared slightly better than approach \#1 in predicting QDAG's complexity. 

\section{Conclusion}
Based on the CHAID model output, we can draw a conclusion that QDAGs which have similar Nodes and edges tend to have similar runtime behaviour. Both approaches discussed have their advantages and disadvantages. If QDAGs operate on a consistent schema, as is the case with on-premise systems, our recommendation would be to implement Approach \#2. In cloud based environments though, where there are several tenants and each tenant operates on different schemas, investing engineering effort in Approach \#1 is worth while, as it provides flexibility in engineering the features in a way that its schema independent. With some fine-tuning and augmentation of runtime statistics of QDAGs, this approach can provide a prediction model that generalizes well across disparate schemas   and is stable in its predictions.

\bibliographystyle{IEEEtran}
\bibliography{IEEEabrv,conference_041818}
\vspace{12pt}
\clearpage

\section{Glossary}
\label{Glossary}
The paper uses few terms related to Spark DAGs. These terms are described here:
\begin{itemize}
\item Schema - A schema defines how data is organized within a relational database; this is inclusive of logical constraints such as table names, fields, data types
\item DAG - Directed Acyclic Graph
\item QDAG - DAG that represents query execution plan 
\item Nodes - Nodes are vertices in the DAG. These represent operators like Scan, Filter, Join
\item Edges - Edge connects two Nodes. Each edge is directed from one Node to another such that following those directions will never form a closed loop
\item Operator Code - Code for Operators in Spark like Sort, Filter, Join, Aggregate 
\item Source Node - Node from which an edge start
\item Target Node - Node at which the edge ends
\item InDegree - InDegree of a node is number of edges coming to the node
\item OutDegree - OutDegree of a node is number of edges coming out of the node
\item Topological(DAG) ordering is a linear ordering of vertices such that for every directed edge uv from vertex u to vertex v, u comes before v in the ordering.
\item Topological Forward Order - Topological ordering of the node starting from root
\item Topological Backward Order - Topological ordering of the node in the reverse order
\item Broadcast Exchange - is a operator in Apache Spark to broadcast rows to other nodes in Apache Spark cluster
\end{itemize}

\end{document}